# Vegetation–climate feedbacks across scales

Diego G. Miralles[1], Jordi Vilà-Guerau de Arellano[2], Tim R. McVicar[3,4], and Miguel D. Mahecha[5]

[1] Hydro-Climate Extremes Lab (H-CEL), Ghent University, Ghent Belgium
[2] Meteorology and Air Quality Group, Wageningen University & Research, Wageningen, Netherlands
[3] CSIRO Environment, Canberra, Australian Capital Territory, Australia
[4] Australian Research Council Centre of Excellence for Climate Extremes, Canberra, Australian Capital Territory, Australia
[5] Institute for Earth System Science and Remote Sensing, Remote Sensing Centre for Earth System Research, Leipzig University, Leipzig, Germany

**Abstract**

Vegetation often understood merely as the result of long-term climate conditions. However, vegetation itself plays a fundamental role in shaping Earth's climate by regulating the energy, water, and biogeochemical cycles across terrestrial landscapes. It exerts influence by altering surface roughness, consuming significant water resources through transpiration and interception, lowering atmospheric $CO_2$ concentration, and controlling net radiation and its partitioning into sensible and latent heat fluxes. This influence propagates through the atmosphere, from microclimate scales to the entire atmospheric boundary layer, subsequently impacting large-scale circulation and the global transport of heat and moisture. Understanding the feedbacks between vegetation and atmosphere across multiple scales is crucial for predicting the influence of land use and cover changes and for accurately representing these processes in climate models. This short review aims to discuss the mechanisms through which vegetation modulates climate across spatial and temporal scales. Particularly, we evaluate the influence of vegetation on circulation patterns, precipitation and temperature, both in terms of trends and extreme events, such as droughts and heatwaves. The main goal is to highlight the state of science and review recent studies that may help advance our collective understanding of vegetation feedbacks and the role they play in climate.

**Introduction**

For centuries, the interplay between climate and vegetation has captivated scientists. It is a relationship of give and take: while vegetation relies on the atmosphere for survival, it also plays a crucial role in shaping Earth's climate. Through a variety of biophysical and biochemical processes, vegetation controls the flow of energy, water, carbon, and other nutrients in the critical zone of our planet. Representative processes include the modulation of wind patterns, moistening of air via transpiration and interception loss, regulation of atmospheric $CO_2$ concentration, and control over surface net radiation and its partitioning into latent and sensible heat fluxes. The study of this interplay can be traced back to Alexander von Humboldt, who theorized that the loss of vegetation could lead to changes in local climate patterns, reducing rainfall and altering ecosystem dynamics (von Humboldt, 1850). His journeys laid the groundwork for substantial scientific advancements, including Vladimir Vernadsky's vision of the biosphere playing an active role in shaping the Earth's biogeochemical cycles (Vernadsky, 1926), and Wladimir Köppen's idea that climatic boundaries are the main drivers of biogeographical patterns (Köppen, 1936). However, it was not until the 1980s, with the development of coupled climate models, which encoded vegetation and soil processes, that the atmospheric community fully embraced the central role of plants in the climate system (Dickinson et al., 1986; Sellers et al., 1986). These models required a mathematical representation of land fluxes regulated by vegetation, which did not conform to the mathematics of fluid dynamics that were used by the atmospheric community. Thereby, even the most traditional atmospheric scientists started to acknowledge the undeniable influence of terrestrial ecosystems on atmospheric processes, and the need to represent plant behavior accurately to predict upcoming weather and future climate (Bonan, 2015).

The impact of vegetation on the climate system is far-reaching, affecting every scale, from local microclimates to global atmospheric circulation, influencing the severity of meteorological extremes, and shaping long-term climate trends (Pörtner et al., 2023; Mahecha et al., 2024; Vilà-Guerau de Arellano et al., 2023). As such, understanding vegetation–atmosphere feedbacks at various spatiotemporal scales is essential to anticipate how the biosphere's adaptation to climate change and land use changes will in turn affect future climate (Bonan et al., 2024). Despite the recognition of its crucial importance since the 1980s, recent reports by the Intergovernmental Panel on Climate Change (IPCC) have only lightly touched upon vegetation feedbacks, with the biophysical ones (e.g., those related to leaf area, vegetation roughness and transpiration) remaining particularly understudied (Forster et al., 2021). This is partly due to a historical focus on atmospheric feedbacks—such as the cloud, lapse rate and water vapor feedbacks—which are, nonetheless, also influenced by vegetation state and activity.

In this short review, we discuss the mechanisms through which vegetation influences climate at various spatiotemporal scales, examining its impact on energy, water, carbon and momentum fluxes, and their myriad of linkages to atmospheric boundary layer thermodynamics, meso-scale and synoptic circulation, and ultimately global precipitation, temperature and humidity patterns. Furthermore, we review the climatic consequences of vegetation changes, with specific emphasis on extreme events, such as droughts and heatwaves. In doing so, the current understanding of vegetation–climate feedbacks is synthesized while highlighting studies that have recently advanced our knowledge on the role that vegetation plays in our climate system.

**Plant's control on energy, water and carbon fluxes**

To review the different pathways by which plants exert control over the state of the atmosphere, the combination of the surface radiation budget and the energy balance equation provides an excellent foundation:

$$R_n = S\downarrow - S\uparrow + L\downarrow - L\uparrow = \lambda E + H + G + [...] \qquad \text{(eq. 1)}$$

where $R_n$ represents surface net radiation, $S\downarrow$ and $S\uparrow$ are the incoming and outgoing shortwave radiation, respectively, and $L\downarrow$ and $L\uparrow$ are the longwave counterparts, which depend on atmospheric and land surface temperature, respectively, following the Stephan Boltzmann law. $\lambda E$ is the latent heat flux associated with evaporation, $H$ the sensible heat flux, and $G$ the ground heat flux. $\lambda E$ and $H$ are turbulent fluxes that depend on gradients between surface and atmospheric properties, while $G$ is controlled by the vertical gradient of temperature (and moisture) in the soil. All these fluxes are typically expressed in units of $W\ m^{-2}$, and more complex versions of this equation include terms such as advective energy and the energy associated with temporal changes in net ecosystem exchange of carbon (i.e., the balance between photosynthesis and respiration) (Jacobs et al., 2008).

A key influence of vegetation on climate comes from the degree to which the plant concentration of pigments and its structural properties, such as the leaf area index (LAI), modify the reflectivity or albedo of the land surface, and therefore how much of the incoming (direct and diffuse) shortwave radiation ($S\downarrow$) is absorbed (and contributes to $R_n$) versus how much is reflected ($S\uparrow$). The albedo of ecosystems is dynamic, varying in both time and space. Ecosystems with high albedo include snow-covered landscapes or deserts, whereas forests typically have a low albedo and thus absorb more energy (Fig. 1). Besides those related to solar incidence angles, temporal changes in surface albedo reflect phenological and disturbance dynamics, ecosystem transformations due to land use change, and ecological succession. These processes thus play a role in regional energy balances and the climate system as a whole. Moreover, the temperature of an ecosystem, and therefore its emission of longwave (L↑), is directly affected by the thermal properties of vegetation, its albedo, and its evaporation rate (λE). On average, vegetated ecosystems warm at lower rates during the day, and thus their L↑ tends to be lower and inversely related to vegetation density (Fig. 1).

Vegetation's control upon $R_n$ is not limited to outgoing radiative fluxes (S↑, L↑); plants also influence incoming radiation (S↓, L↓) through complex interactions with the atmospheric boundary layer (ABL) that control convective cloud formation (see below). For instance, via emission of biogenic volatile organic compounds (BVOCs) that act as aerosol precursors and can form cloud droplets, and through the plants' regulation of atmospheric $CO_2$ concentration (Fig. 1). By absorbing $CO_2$ during photosynthesis, which is then released over various time scales through respiration, ecosystems modulate the greenhouse effect of our planet and thus L↓. Under certain conditions, methane ($CH_4$) can represent the final stage of terrestrial carbon cycling in ecosystems. This exchange plays a vital role in the Earth's greenhouse gas budget, thereby influencing the global energy balance, temperature and long-term climate patterns, as demonstrated in paleoclimatic records (Canadell et al., 2021). In fact, the enhanced photosynthesis and subsequent greening that followed from $CO_2$ fertilization and global warming in recent decades has led to the sinking of approximately one third of anthropogenic $CO_2$ emissions (Friedlingstein et al., 2023). The consequent dampening of the secondary greenhouse effect in our planet implies a negative biochemical feedback on temperature that has been quantified as approximately –0.8 W m$^{-2}$ C (Canadell et al., 2021). In addition to $CO_2$, $CH_4$ and water vapor ($H_2O$), vegetation also influences the concentration of other potent greenhouse gasses, such as nitrous oxide ($N_2O$), and ozone ($O_3$) (Stocker et al., 2013).

Plants do not only affect all four radiative fluxes in eq. (1), but also the partitioning of net radiation ($R_n$) among λE, H and G. By reducing land surface temperature and the land's thermal conductance, vegetated ecosystems typically experience lower G than their surroundings (Bonan, 2015). This lower G is also a consequence of how the physical structure and density of vegetation enhance the tendency of the near-surface atmosphere to conduct and convect mass and heat, favoring the dissipation of $R_n$ into the atmosphere via turbulent fluxes (λE and H). In dense forests with tall trees, surface roughness creates a drag on the airflow, reducing average wind speed but enhancing turbulence within and just above the canopy. The net effect is increased aerodynamic conductance and thus enhanced exchange of momentum, heat, and mass from land to atmosphere (Fig. 1). In fact, recent studies show the key relevance of accurately representing aerodynamic conductance within the canopy layer in climate models (Bonan et al., 2021), and attribute the influence of greening on climate mostly to increases in aerodynamic conductance (Chen et al., 2020).

The partitioning of available energy (AE = $R_n$ – G) between λE and H is also dynamically controlled by vegetation and varies spatio-temporally across ecosystems. In forested areas, when sufficient soil moisture is available and energy-limited conditions prevail, a higher proportion of AE is converted into λE, leading to a cooling effect on local climate (Fig. 1). This phenomenon is attributed to the extensive leaf area and deep roots, leading to a high water uptake and loss through transpiration, but also to the intense flux of interception loss during and after periods of rain. The water intercepted by plants and subsequently evaporated without reaching the soil can have a crucial importance for humidity, fog and cloud formation, and has complex implications for the energy balance of forested ecosystems (van Dijk et al., 2015). Conversely, regions characterized by sparse vegetation and low transpiration, such as drylands, predominantly channel net radiation into sensible heat flux, which contributes to warmer atmospheric conditions. This dynamic partitioning of AE between the two turbulent fluxes affects local weather patterns, atmospheric stability, and precipitation processes, and has been at the core of the study of ecosystem's influence on the atmosphere for at least a century (Bowen, 1926).

Ultimately, plants act as a nexus between the energy, carbon, and water cycles. Through transpiration and interception loss, vegetation links energy and water cycles, not only cooling but also moistening the surface layer of the atmosphere and affecting humidity and cloud formation in the ABL (see next section). Likewise, vegetation connects energy and carbon cycles through the consumption of photosynthetically active radiation (PAR) during photosynthesis, and through the intake and emission of $CO_2$ that controls $L\downarrow$. Finally, photosynthesis and transpiration—and therefore carbon, water and energy cycles—are intrinsically linked through the stomata openings on the leaf, which regulate the exchange of $CO_2$, water vapor, and oxygen with the atmosphere (Meidner and Mansfield, 1968). Stomatal conductance is influenced by environmental factors such as photosynthetically active radiation (PAR), temperature, and vapor pressure deficit (VPD), that also depend on vegetation–atmosphere feedbacks, and its optimization reflects the delicate balance plants maintain to maximize carbon gain while preventing dehydration (Medlyn et al., 2011). This interconnection of cycles ensures a cascading effect from one to another, emphasizing the need for holistic approaches to understand how plants, and the ecosystems they co-create, shape atmospheric conditions across scales (Vilà-Guerau de Arellano et al., 2023).

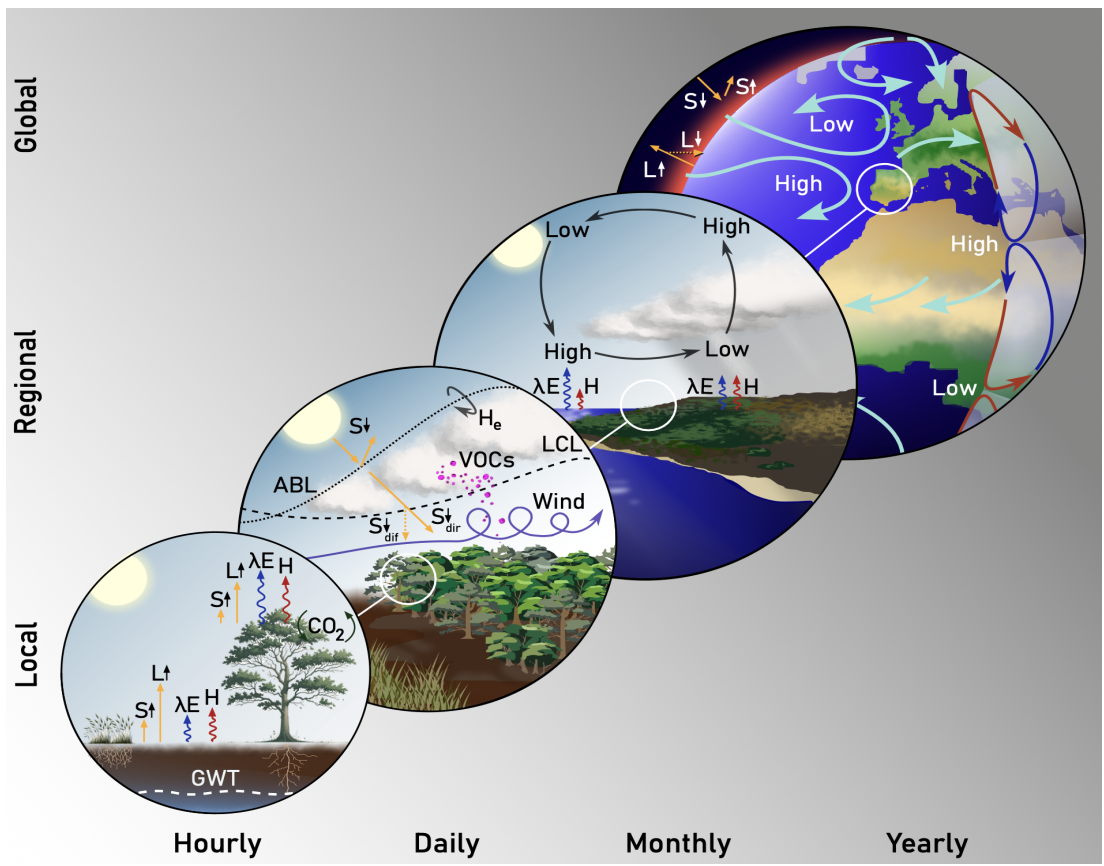

**Figure 1**. Influence of vegetation on the atmosphere across meteorological scales. Vegetation controls the surface energy, water and carbon fluxes at local scales, influenced by soil moisture (and the dynamics of the groundwater table, GWT). This influence propagates to the atmosphere via turbulent fluxes, causing convective and mechanical instability, altering the state and diurnal growth of the ABL, regulating moisture and heat entrainment ($H_e$), and thus the lifting condensation level (LCL) and convective cloud formation. At larger scales, vegetation influences meso-scale circulation and the location of semi-permanent low and high pressure systems, affecting the redistribution of heat, water and carbon, and thus influencing temperature, humidity and precipitation patterns at planetary scales.

**Local atmosphere response to vegetation**

The influence of vegetation on the local atmosphere extends from microscale interactions at the stomata to ABL dynamics. The biophysical and biochemical processes described in the previous section influence the humidity and temperature diurnal variability and profiles, ABL growth, and atmospheric thermodynamic stability. A key process governing the exchange of heat, water and carbon in the land–atmosphere interface is turbulence. Turbulence can either be mechanically generated by wind instabilities or triggered by air density (temperature) gradients, with both being directly controlled by vegetation's presence, structure, state and activity, and influenced by biodiversity—a term that is understood today as being broader than just species richness. Through its influence on roughness and turbulent fluxes, vegetation affects the formation of convective clouds and rainfall, and therefore also the partitioning of S↓ between diffuse and direct radiation, which is crucial for photosynthesis itself (Fig. 1).

The study of the impact of the land surface on the ABL can be traced back to the mid 20th century (Monin and Obukhov, 1954), yet the field gained significant stimulus in the latter half, leveraging advances in computer and data sciences, instrumentation (e.g., eddy covariance), and theoretical understanding. By the 1970s and 1980s, the advent of more sophisticated land-surface schemes in numerical weather prediction (and observational techniques, such as remote sensing) allowed for a more detailed examination of vegetation influences on ABL (thermo)dynamics. Large Eddy Simulation (LES) models also emerged as a powerful computational tool to study explicitly turbulent flows (Deardorff, 1970). However, it was not until the late 20th century that LES was employed for the first time to study ABL dynamics and their dependence on terrestrial ecosystems (Albertson and Parlange, 1999).

Today we understand that the stability and growth of the ABL depends on boundary conditions such as subsidence, advection, and the properties of the residual and free tropospheric layers. However, the diurnal development of the ABL is first and foremost triggered by surface turbulent fluxes (regulated by vegetation) and their influence on the entrainment of air from the residual layer and free troposphere as the ABL grows. In dry conditions and bare (or sparsely vegetated) land, limited transpiration results in higher H and a warming of the near-surface layer. This often creates a strong thermal instability that promotes the rapid growth of the ABL. Nonetheless, the absence of moisture often yields a drier and warmer ABL, which can be associated with reduced cloud formation unless moisture is entrained or advected from elsewhere (Taylor et al., 2012). Conversely, the presence of dense and active vegetation allows for enhanced transpiration and interception loss, contributing to cooling the surface and adding moisture to the air (Fig. 1). A reduced H may favor the development of more stable, moist convective boundary layers, which tend to grow more slowly than those over bare soils. However, the lower albedo, higher moisture content, and strong mechanical turbulence may still lead to preferential cloud formation over forested regions compared to their surroundings (Teuling et al., 2017).

The diurnal growth of the ABL is ultimately enabled by the entrainment of air from the free troposphere (Garratt, 1994). Entrained air masses, with residual characteristics from previous days or free tropospheric conditions, influence the state of the ABL and the near-surface atmosphere (Fig. 1). Moreover, the advection of air masses (due to sea breeze or synoptic systems) also exerts an influence on both ABL development and vegetation activity. In the morning, the air that is entrained as the ABL grows is comparatively warm and dry (van Heerwaarden et al., 2009) and has a low $CO_2$ content (Vilà-Guerau de Arellano et al., 2012). This air enhances VPD at the canopy level, and leads to increased evaporation, thus shifting the turbulent heat flux partitioning. This effect can be partly offset by the closure of stomata to regulate plant transpiration under limited water availability. Likewise, the entrainment of air with limited $CO_2$ concentration can be a key factor in down-regulating photosynthesis during the day and affecting stomatal conductance (Dupont et al., 2024; Vilà-Guerau de Arellano et al., 2012). This dynamic interplay between vegetation and ABL during the day is crucial for the formation and intensification of boundary-layer clouds, such as shallow cumulus (Pedruzo-Bagazgoitia et al., 2019; Vilà-Guerau de Arellano et al., 2023). Moreover, modifications of vegetation activity by the combined effects of cloud shading, and changes in T, $CO_2$ concentration and VPD lead to shifts in the partitioning between H and λE, which in turn influence turbulent transport of heat and moisture in the ABL. This also changes cloud coverage, cloud microphysics, and the capacity to move air masses from the ABL into the free troposphere (Sikma and Vilà-Guerau de Arellano, 2019).

Finally, BVOCs emitted by vegetation also play a critical role in influencing radiation, temperature, and local precipitation patterns (Fig. 1). BVOCs, and other biogenic-origin substances such as pollen, contribute to the formation of aerosols and cloud condensation nuclei, affecting cloud properties, the partitioning of direct and diffuse incoming shortwave radiation (including PAR), and the Earth's atmosphere radiative balance (Durand et al., 2021). These compounds may lead to the cooling of the atmosphere by increasing cloud reflectivity, yet they may also cause warming, since they enhance the lifetime of methane and contribute to the formation of ozone and other greenhouse gasses in the presence of nitrogen oxides (Sporre et al., 2019). The impact of BVOCs on cloud formation and temperature is complex and varies depending on the relative concentrations of specific BVOC molecules and the background atmospheric chemistry. Field research shows that in boreal forest under stress, BVOCs increase promotes cloud formation (Joutsensaari et al., 2015), and that certain pollen types can have analogous influences on clouds (Casans et al., 2023).

In summary, vegetation regulates the amount of moisture available for convective cloud formation through transpiration and interception loss. Cloud formation also depends on turbulence, which transports heat and moisture and yields ABL growth, and is also directly controlled by the state of the ecosystem. Furthermore, vegetation emits BVOCs and pollen that enhance condensation nuclei and can be key in the formation of cloud and rain droplets. Clouds and aerosols, in turn, regulate the amount of radiation and the fraction of diffuse PAR reaching the surface, further illustrating the complex feedback loops between vegetation, the ABL, and cloud formation. Predicting upcoming weather as well as future climate conditions requires accurate understanding of this two-directional vegetation–atmosphere coupling. In particular, it remains unclear how these bidirectional effects mutually interact during extreme conditions (Mahecha et al. 2024). Recent LES simulations have enabled the exploration of different future scenarios of $CO_2$ fertilization and warming—revealing changes in photosynthesis and ABL conditions for different scenarios—and have highlighted the need for integrated studies that consider soil, canopy, and atmospheric properties holistically (Sikma and Vilà-Guerau de Arellano, 2019). This underscores the importance of comprehensive field campaigns to constrain and evaluate dedicated numerical simulations that explicitly resolve these two-directional interactions and advance our understanding of this coupled system (Bonan et al., 2021, 2024; Vilà-Guerau de Arellano, 2024).

**Vegetation influence on atmospheric dynamics**

Vegetation also exerts a significant influence on atmospheric dynamics, impacting wind patterns at a wide range of spatial scales (Fig. 1). In fact, vegetation has been proposed as a key factor in processes such as the slowdown of global near-surface winds (Vautard et al., 2011), moisture convergence over forests (Makarieva and Gorshkov, 2007), and expansion of the Hadley cells (Shin et al., 2012). Furthermore, it is known to impact meso-scale circulation by modifying thermal and moisture contrasts with surrounding regions (Fig. 1). Vegetation may enhance or dampen sea breezes and valley–mountain flows (Mostamandi et al., 2022), and may even influence monsoon intensity (Cui et al., 2020). Moreover, semi-permanent highs and lows, which are critical features in our Earth's climate system, are also influenced by the presence of land, and likely modulated by changes in land cover and vegetation (McPherson, 2016); the distribution and characteristics of vegetation affect the albedo, moisture availability, and thermal properties of the land, which are in turn expected to influence synoptic-scale atmospheric pressure patterns (Fig. 1). For example, large forested areas can increase near-surface atmospheric temperature and humidity, potentially weakening (strengthening) high-pressure (low-pressure) systems (McPherson, 2016). This influence of vegetation on local-to-global atmospheric dynamics underscores the critical role of terrestrial ecosystems in global climate regulation, which ultimately contributes to the correlation between global ecosystem distribution and climate patterns (Köppen 1936).

The term 'global stilling' refers to the observed reduction in terrestrial near-surface wind speeds measured in recent decades over land (Roderick et al., 2007). This phenomenon contrasts with the expected increase in wind activity in a warming world, and with the observed increasing trends in wind speed over the oceans (Young et al., 2011). Increasing near-surface wind speeds at higher latitudes has been reported in both hemispheres, pointing to important regional variations over land (McVicar et al., 2012). These changes have been linked to increases in surface roughness, primarily due to vegetation growth (Vautard et al., 2011). Nonetheless, subsequent analysis based on near-surface wind speeds observations along with a conceptual boundary layer model attributed wind speed changes to changes in roughness, but the precise drivers, such as urbanization or forestation, were less clearly defined (Wever, 2012). In fact, later work using Earth System Models to isolate the response of near-surface wind speed to increases in LAI, found that enhanced LAI was not a dominant driver of global stilling (Zeng et al. 2018a). Finally, it should be noted that the rate of stilling has seemingly weakened or even reversed in recent decades (Zeng et al., 2019). These insights highlight the complex, yet uncertain, role vegetation plays in shaping local and regional wind patterns, underlining the need for enhanced observational capabilities and modeling efforts to understand the drivers behind changes in near-surface wind speeds.

At the planetary-scale, the Hadley circulation influences weather patterns, including the positioning and intensity of storm tracks, subtropical high-pressure systems, jet streams, and tropical monsoons. Multiple studies have reported a poleward expansion of the Hadley cells as our climate warms (Marvel and Bonfils, 2013; Seidel et al., 2008), some of them relating this expansion to land–atmosphere feedbacks in drylands (Shin et al., 2012; Song and Zhang, 2007). As drylands expand and vegetation diminishes, the resulting increase in surface albedo creates a feedback mechanism that may help expand the Hadley cells poleward. These vegetation-driven changes of global circulation were first postulated by Charney (1975), who hypothesized that a reduction of vegetation and consequent increase in albedo in the Sahel region would intensify the sinking of the Northern Hemisphere Hadley cell and perpetuate arid conditions. The expansion of the Hadley cells has important implications for regional water availability and has already been linked to drought intensification in regions like Australia (Post et al., 2014). The influence of vegetation on these dynamics, particularly through changes in albedo, highlights again the complex interactions between terrestrial ecosystems and global atmospheric patterns. Understanding these vegetation-driven changes in (sub)tropical circulation is crucial to accurately predicting and managing the role of drylands in their own expansion (Koppa et al., 2024).

Arguably the most controversial thesis regarding the role of vegetation on global circulation is the 'biotic pump theory' postulated by Makarieva and Gorshkov (2007). The theory focuses on the importance of condensation-induced atmospheric dynamics, positing that the large transpiration from forests, and the subsequent condensation over them, lowers the water vapor pressure in the lower atmosphere and leads to increased convergence of moisture from surrounding areas. Indeed, condensation affects atmospheric pressure through both latent heating and water vapor mass removal; while it is commonly accepted that the increased pressure due to latent heating dominates, Makarieva and Gorshkov (2007) defended the important role of water vapor mass removal for atmospheric dynamics. As such, the theory implies that forests exert a profound influence on regional and global weather patterns by substantially enhancing moisture transport from oceans to land. This 'moisture pull' of forests results in increased precipitation over terrestrial areas, and it also stabilizes and extends rainfall patterns. This theory has been heavily contested, yet seemingly without definitive resolution (Jaramillo et al., 2018; Meesters et al., 2009). Given the increased disturbance of forest ecosystems, understanding the mechanisms behind the biotic pump is critical for predicting changes in global weather patterns and developing strategies to mitigate the adverse effects of deforestation (or afforestation and reforestation) on climate system dynamics.

In summary, the complex interplay between vegetation and atmospheric dynamics extends beyond local turbulence within the atmospheric boundary layer, influencing major atmospheric weather processes and global climate patterns. The influence at meso-scales is potentially important for regulating sea breezes and even monsoonal circulation. At a larger scale vegetation may influence, for example, the subsidence associated with the Hadley cell circulation and the location of semi-permanent atmospheric pressure patterns. The global influence of vegetation is also seen in phenomena such as global stilling, with increased surface roughness due to vegetation growth affecting near-surface wind speeds. Moreover, the extensive transpiration of large forested areas can seemingly enhance moisture transport from oceans to land, stabilizing regional climates and modifying rainfall patterns. These dynamic interactions highlight the role of vegetation in climate regulation. Advanced modeling and comprehensive observational strategies are required to fully understand their importance and to predict their implications in future climates.

**Vegetation feedbacks and climate trends**

Understanding the processes by which vegetation influences the atmosphere across spatial scales is only a first, yet necessary, step to understand how biophysical and biochemical feedbacks will shape temperature and precipitation as we move into the future. Climate perturbations associated with greenhouse gas (and aerosol) emissions and land use forcing have an influence on vegetation that spans from minutes to seasons and to millennia. This influence is in turn expected to either dampen (negative feedback) or amplify (positive feedback) the initial climate perturbations. Observational studies show that the recent tendency towards $CO_2$-richer and warmer atmospheres has already led to global greening in recent decades (Piao et al., 2020), largely due to earlier growing seasons (Lian et al., 2022) (Fig. 2). Moreover, the imprint of global greening on recent trends in temperature (Forzieri et al., 2017) and precipitation (Zeng et al., 2018b) have already been documented. Yet, this global greening trend has been showing signs of deceleration due to nutrient and water limitation, among other factors (Peñuelas et al., 2017; Winkler et al. 2021), and there is a risk that the greening trend is even reversed in future climates (Zhang et al. 2022). In addition to global greening, the emission of $CO_2$ also had repercussions for vegetation's phenology, influencing the senescence of leaves, altering surface albedo, the evaporation of water through transpiration and interception loss, the roughness of the ecosystem, and the entire carbon cycle (Lian et al., 2022). The impact of these phenological shifts on the climate system, particularly on precipitation and runoff patterns but also on temperature, remains an area of active research (Piao et al., 2020).

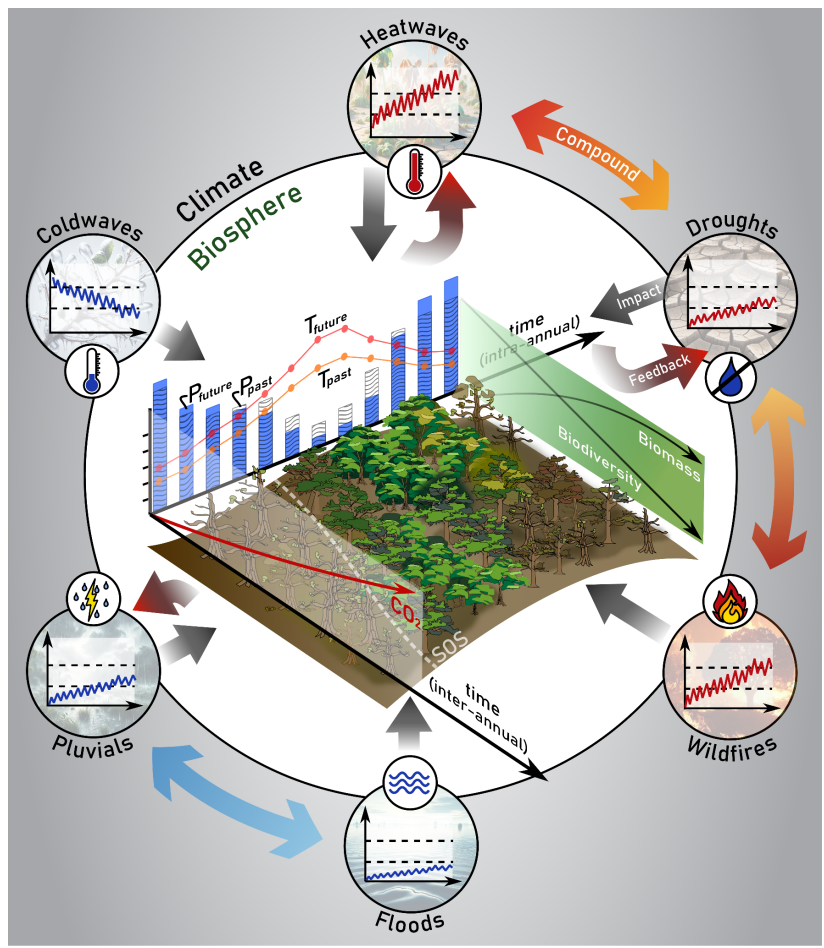

**Figure 2**. Influence of vegetation on climate trends and extremes. Ecosystem trends, including the tendency towards larger biomass and lower biodiversity, succession and acclimation, or phenological trends—such as the earlier start of the spring (SOS) —are expected to influence temperature and precipitation trends. Moreover, extreme climatic events do not only influence the ecosystems but may also be affected by ecosystem dynamics and vegetation structure and activity through multiple positive and negative feedbacks

In addition to global greening and phenological changes, climate change is also expected to affect ecological succession by changing the ecological niche (Antão et al. 2022), favoring alien species that may transform autochthonous communities (Essl et al. 2020). Likewise, climate change can trigger large-scale tree mortality due to a combination of plant stress and more favorable conditions for fungi or beetle infestations (Netherer et al. 2024). As a result of changes in species composition, ecosystem structure and plant functional traits (e.g., root depth, leaf mass, conduit density, leaf nitrogen and phosphorus) may also change (Diaz et al. 2016). Ecologists have argued that changes in species richness and functional diversity (i.e., the diversity of plant traits) have an imprint on the functioning of ecosystems as a whole (Reichstein et al. 2014, Musavi et al. 2015). And in fact, it has been shown that the prevalent plant functional traits and their diversity are related to ecosystem-scale functional properties such as carbon uptake potential and water/light use efficiency (Migliavaca et al. 2021). Changes in vegetation composition may occur progressively or abruptly as a tipping point (Bonan 2015), and the subsequent functional diversity can be very different. In any of these cases, multi-temporal changes in terrestrial ecosystems are expected to feedback on regional, and most certainly global, long-term climate trends (Willeit et al., 2014).

According to the recent IPCC AR6 report (Canadell et al. 2021), and up-to-date $CO_2$ assessments (Friedlingstein et al., 2023), the enhanced photosynthesis that has followed $CO_2$ fertilization in recent decades has led to the sinking of approximately one third of anthropogenic $CO_2$ emissions over land. The consequent dampening of the greenhouse effect implied a negative biochemical feedback on temperature of around $-0.8$ W m$^{-2}$ C$^{-1}$, yet with an uncertainty that is almost an order of magnitude higher over land than over the ocean (Arora et al., 2020). Should this be the magnitude of the feedback, it would equate to a buffering of global warming by around $-0.5$ °C since pre-industrial times; to give some perspective, that is around half of the warming induced by the water vapor feedback, the strongest positive feedback in nature. However, around one third of the cooling associated with this $CO_2$ biochemical feedback is thought to have been compensated by the detrimental influence that the climate response to $CO_2$ emissions had on photosynthesis (Canadell et al. 2021). This mainly relates to $CO_2$-driven trends in soil moisture and temperature reducing gross primary production (Huang et al., 2019; Stocker et al., 2019). Finally, non-$CO_2$ related biogeochemical feedbacks (such as those associated with the influence of climate on BVOC emissions, or the land release of $CH_4$ and $N_2O$) are reported by the IPCC AR6 as in the order of $-0.16$ W m$^{-2}$ C$^{-1}$, yet with a high uncertainty (Canadell et al. 2021). The influence of these biochemical responses on evaporation, precipitation and runoff is even more uncertain, and remains an outstanding research gap (Yang et al., 2019).

Despite a historical focus on physical feedbacks—such as the snow albedo and atmospheric feedbacks (e.g., cloud, lapse rate, water vapor feedbacks)—the IPCC has traditionally concentrated preferentially on biochemical rather than biophysical feedbacks. Feedbacks associated with changes in leaf area, roughness and/or soil moisture controls on transpiration, have only been lightly touched upon by IPCC reports, likely due to the more limited number of studies and large uncertainties. Therefore, the influence of these biophysical feedbacks on climate trends remains a crucial research gap. Current estimates of the net biophysical feedback range from close to zero (Willeit et al., 2014) to $+0.13$ W m$^{-2}$ C$^{-1}$ (Stocker et al., 2013), while paleoclimatic approaches point to larger estimates, around $+0.3$ W m$^{-2}$ C$^{-1}$ (Forster et al. 2021). Given this limited evidence and high divergence among existing studies, the recent IPCC AR6 estimated the biophysical feedback as $+0.15 \pm 0.15$ W m$^{-2}$ C$^{-1}$, assigning it a low confidence (Forster et al. 2021). And once again, the influence that changes in biophysical properties may have on the water cycle as we progress into the future remains even more uncertain (Yang et al. 2019). Finally, the influence of global greening on the surface albedo feedback—which is, nonetheless, mostly dominated by snow and sea ice variability—is thought to be relatively limited (Forster et al. 2021), yet several studies have reported a warming associated with a shift from tundra to boreal forests in Northern Hemisphere high latitudes (Willeit et al., 2014) (Armstrong et al., 2019).

In addition to vegetation–climate feedbacks, the forcing associated with direct human perturbations—such as clearing land for agriculture, reforestation of abandoned farmland, and urbanization—has a direct and long-lasting impact on terrestrial ecosystems and our climate system. In fact, the overall effects of anthropogenic land-use and land-cover changes may be comparable in magnitude to climate-induced vegetation changes (Davies-Barnard et al., 2015). Over decades to centuries, these land use changes drive successional shifts that alter community composition, ecosystem structure, surface energy fluxes, soil properties, carbon storage, and greenhouse gas emissions, thereby influencing trends in temperature (Pongratz et al., 2021) and precipitation (Hertog et al., 2024). Moreover, land cover changes have been highlighted as potential drivers of wind stilling over land (Vautard et al., 2011), the expansion of the Hadley cells (Song and Zhang, 2007), and even of the slow-down of the Atlantic Meridional Overturning Circulation (Armstrong et al., 2019). Nonetheless, the biophysical (and biochemical) processes associated with long-term ecological shifts and phenological changes are still not dynamically represented in many global climate models, which may lead to inadequate climate model projections of the influence of vegetation on hydrology and climate trends.

The forcing associated with land cover change and its influence on albedo has recently been estimated as $-0.15$ W m$^{-2}$ since 1700 and $-0.12$ W m$^{-2}$ since 1850, and likely resulted in a net global cooling of about 0.1 °C since 1750 (Eyring et al. 2021). Moreover, the IPCC Special Report on Climate Change and Land assessed that there is robust evidence and high agreement that land cover and land use or management exert important influence on temperature, rainfall and wind intensity at various spatial and temporal scales, through biophysical feedbacks (Jia et al., 2019). In light of this important influence, the intentional climate modification through land geoengineering strategies—such as reforestation or changes in land use—offers potential pathways for climate mitigation and adaptation (Seneviratne et al., 2018b). Such strategies should aim at exploiting vegetation's natural capabilities to cool the local environment and/or enhance precipitation, thereby counteracting some of the negative effects of climate change.

**Vegetation feedbacks during hydro-climatic extremes**

As seen above, understanding the role of vegetation feedbacks in shaping long-term climate trends over multiple spatial scales is crucial. However, understanding the dynamic influence of vegetation feedbacks at the scale of extreme climatic events is even more relevant if we aim to mitigate their societal and ecosystem consequences (Mahecha et al. 2022, 2024). Extreme events—such as droughts, heatwaves, coldwaves, wildfires, storms and floods—directly affect water availability, agricultural productivity, ecosystem services, and human wellbeing (Miralles et al., 2019). Their regional exacerbation and more frequent concurrence as 'compound events' (Zscheischler et al. 2020) are already felt around the world, highlighting the urgent need to understand their drivers for climate adaptation and resilience strategies (Seneviratne et al., 2018a). Since ecosystems are severely affected by climate events, and dynamic changes in vegetation state and activity influence the surface energy balance (see above), vegetation–climate feedbacks are expected during these events (Miralles et al., 2019). Figure 3 provides an overview of the impact of vegetation disturbances on the surface energy balance, by illustrating the anomalies in $\lambda E$ and $H$ during times in which LAI anomalies drop below their 10th percentile (computed per pixel, considering the 1981–2023 period). Overall, lower-than-usual $\lambda E$ and higher-than-usual $H$ can be observed when vegetation is in stressed conditions, and this is particularly the case for water-limited regions where hydro-climatic variability is larger. In high latitudes, low LAI events are often related to low radiation conditions, which lead to anomalously low values of both $\lambda E$ and $H$, while signals are more confounded in tropical forests where variability is low and data tend to be uncertain.

While the influence of vegetation on precipitation volumes and moisture recycling has been studied for decades (Eltahir and Bras, 1996), its influence on the occurrence of pluvials (i.e., periods of excessive rainfall leading to abnormally wet conditions) has seldom been studied. Nonetheless, a recent study demonstrated that more than half of the extreme rainfall during the 2021 European summer storms originated from plant transpiration and interception loss (Insua-Costa et al., 2022). In other words, while vegetation plays a crucial role reducing overland flow and the risk of fluvial floods—by increasing the soil infiltration capacity, preventing erosion and reducing sediment load in water bodies—it can also exacerbate storms by increasing atmospheric moisture content, triggering convection, and inducing meso-scale circulation patterns (see previous sections). Nonetheless, a recent modeling experiment concluded that afforestation in Europe decreases both the number and intensity of extratropical cyclones due to the increased surface roughness, even if convective summer storms are enhanced by afforestation due to increased transpiration (Belušić et al., 2019). Overall, the influence of vegetation on flood occurrence due to its control upon precipitation intensity needs to be considered in integrated assessments of land cover management aiming to reduce the risk of flood events, particularly in coastal regions where fluvial and pluvial floods are expected to be increasingly compounded with storm surges (Ward et al., 2018).

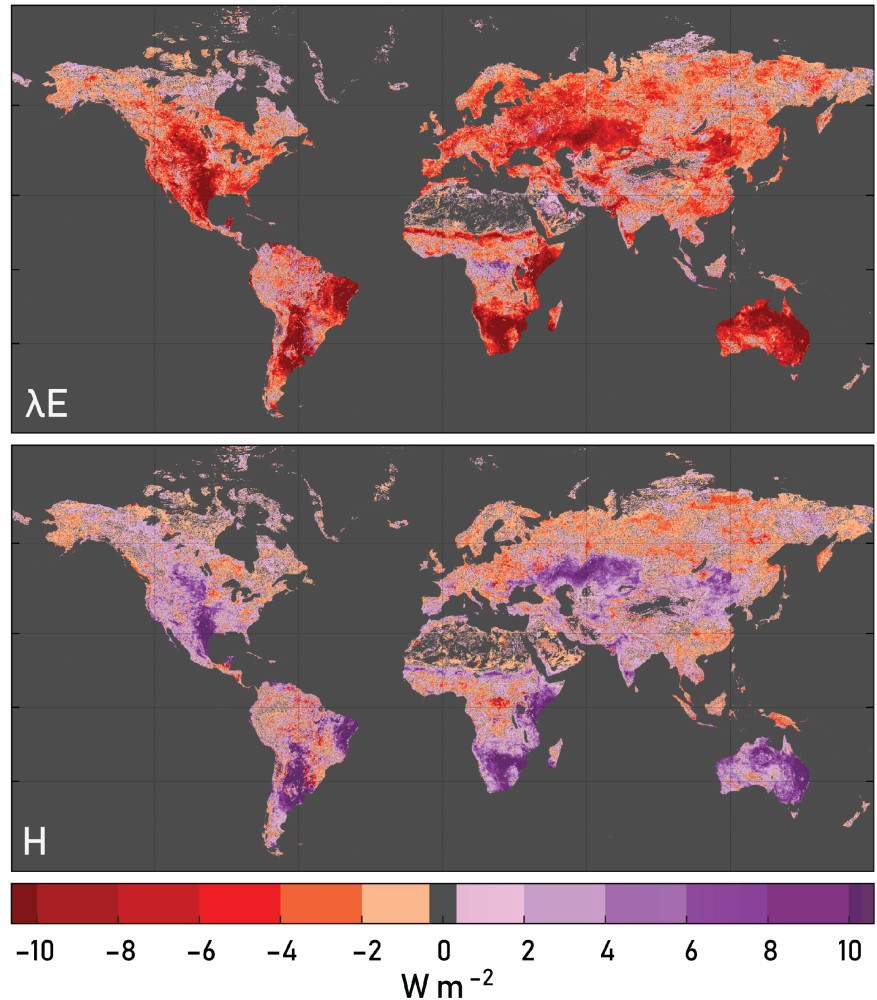

**Figure 3.** Impact of vegetation disturbances on the surface energy balance. Anomalies in λE and H (with respect to their local monthly climatology) during times in which LAI anomalies drop below their 10th percentile (also computed based on the pixel LAI climatology). The 1981–2023 period is considered. Data of λE and H come from GLEAM4 (https://www.gleam.eu) while LAI data come from GLOBMAP (https://zenodo.org/records/12698637).

Droughts and heatwaves are expected to aggravate and synchronize as we progress into the future (Orth et al., 2022; Fig. 2). Current understanding suggests that similar persistent large-scale anticyclonic conditions (i.e., blocking highs) are behind the triggering of both events, while analogous land–atmosphere feedbacks—particularly through vegetation and soil moisture dynamics—play a crucial role in their intensification and propagation (Barriopedro et al., 2023). In general, reduced evaporation from drying vegetation and soils leads to an increase in sensible heat, potentially leading to heatwave escalation, while also diminishing rainfall likelihood and further intensifying drought. In general, vegetation can modulate meteorological drought conditions through the moistening and warming of the atmosphere via transpiration, influencing local and regional humidity, convective stability, and precipitation. Healthy vegetation may mitigate meteorological drought by maintaining a certain level of moisture cycling within the ecosystem (Miralles et al., 2016). Conversely, reduced vegetation activity as soils desiccate may lead to decreased transpiration, potentially exacerbating meteorological drought. In that sense, meteorological droughts can self-intensify via land feedbacks (Schumacher et al., 2022). This is not only a local process, since the advection from dry ecosystems reduces air humidity and precipitation efficiency downwind, providing a mechanism for drought self-propagation that can be determinant in semiarid regions (Schumacher et al., 2022).

During heatwaves, vegetation tends to cool the land surface by providing shade, plus transpiration consumes energy that would otherwise be available for sensible heating and warming of the environment. As such, urban green spaces, including parks and green roofs, have been shown to reduce the urban heat island, making cities more resilient to heatwaves (Barriopedro et al., 2023). Nonetheless, during the first phases of a heatwave, forested ecosystems can be substantially warmer than the surrounding due to their low albedo and more conservative use of water; then as the event progresses, a sustained level of transpiration, enabled by this conservative use of water and their deeper roots, tends to lead to cooling compared to surrounding ecosystems (Teuling et al., 2010). In other words, soil moisture–temperature feedbacks tend to be more positive during the onset, and less positive during the peak of heatwave events in forested areas. Moreover, these vegetation–climate feedbacks can lead again to teleconnected impacts, once downwind advection causes the spatial propagation of the event (Schumacher et al., 2019). Finally, the drying of vegetation combined with dry and hot atmospheric conditions has shown to enhance the risk of wildfires during compound dry–hot events, a situation that is expected to aggravate in the future (Fan et al., 2023). Needless to say that wildfire emissions affect temperature, clouds, and rain, through the emission of greenhouse gasses and aerosols and changes in atmospheric stability (Liu et al., 2014).

Changes in plant phenology also feedback on the occurrence of extreme climatic events. For instance, studies of European heatwaves show that delayed or weak growing season green-up can amplify extreme heatwaves (Lorenz et al., 2013). In contrast, early and strong green-up contributes to enhanced transpiration and surface cooling, thereby initially decreasing the magnitude of warm temperature anomalies (Stéfanon et al., 2012). However, at seasonal scales some of these influences can be complex and seemingly counterintuitive. An earlier and more intense growing season, due to higher temperatures, can yield higher spring transpiration and lead to drier soils in summer, even without anomalies in precipitation (Lemordant et al., 2016). This has the potential to amplify the impact of heat/droughts events later in the year due to a depleted soil moisture state (Sippel et al. 2017). Meanwhile, in addition to the importance of biodiversity for ecosystem resilience during climate extremes, its role in dampening the occurrence of certain extremes has also recently been highlighted (Mahecha et al., 2022, 2024). Biodiversity— when understood in its broadest sense i.e. including genetic, functional, structural diversity, and landscape diversity—can influence the capacity of ecosystems to buffer against climate extremes, affecting processes like carbon sequestration, water retention, and overall ecosystem productivity. For instance, a higher functional diversity can enhance resilience to extreme conditions by providing a range of responses to stress and disturbance ('insurance hypothesis', Loreau et al. 2021). In the context of droughts and heatwaves, ecosystems with higher biodiversity may be better equipped to maintain function (e.g., transpiration, photosynthesis) due to the presence of species that can tolerate a range of conditions. This, in turn, can moderate local climate conditions and feed back into larger climate system dynamics (Mahecha et al., 2024). The concept of biodiversity–climate feedbacks emphasizes the critical role of vegetation–atmosphere interactions in the Earth's climate system, and the need to consider ecosystem dynamics in climate adaptation policies (Pörtner et al. 2023, Bonan et al. 2024).

Despite technical advances and discoveries, critical research challenges remain in disentangling vegetation's role during drought and heatwave events. Operational forecasts and climate models still struggle to capture the complexity of these vegetation–atmosphere interactions, leading to inaccuracies in early-warning systems and climate extreme projections. The IPCC AR6 concludes—with medium confidence due to limited studies and evidence—that vegetation changes can amplify or dampen extreme events through changes in albedo and evaporation, influencing future trends in these events; it also concludes that urbanization increases the risks associated with extreme events by suppressing evaporative cooling and amplifying heatwave intensity (Seneviratne et al. 2021). Moreover, the AR6 affirms that there is robust evidence that dry soil moisture anomalies favor summer heatwaves, and that part of the projected increase in heatwaves and droughts can be attributed to these feedbacks (Seneviratne et al. 2021). The acknowledgment that vegetation feedbacks play a role in exacerbating or mitigating droughts and heatwaves paves the way for exploring geo-engineering strategies aimed at modifying land surface conditions to attenuate these extremes. Measures such as altering crops albedo, modifying irrigation practices, implementing afforestation and/or reforestation have been proposed (Seneviratne et al., 2018b; Thiery et al., 2020). However, the effectiveness of such strategies requires a comprehensive understanding of vegetation–climate interactions, highlighting the need for advanced models and observational tools to improve our predictive capabilities if we aim to develop sustainable mitigation approaches.

**Conclusion**

As climate change continues to shape vegetation dynamics, there is an urgent need to explore how these ecosystem changes will in turn affect climate, and ultimately feed back on biodiversity and ecosystem services. This short review highlights the role of vegetation in regulating atmospheric dynamics across all scales, from local effects in the atmospheric boundary layer to impacts on global circulation patterns. The interactions discussed emphasize the importance of accurately representing biological processes and their coupling in climate models. The complexity of vegetation–climate feedbacks requires an interdisciplinary approach that integrates knowledge from biology, ecology, chemistry, hydrology, meteorology, and climatology. As we continue to gather experimental evidence for changes in vegetation dynamics due to climate change and human activities, it becomes increasingly important to refine our models and strategies based on robust interdisciplinary research. This will enable us to better predict climate trends, prepare for future changes, and implement effective mitigation and adaptation strategies that leverage the natural regulatory capabilities of vegetation.

Several research gaps remain, including the role of vegetation in cloud formation and precipitation, the quantification of individual biophysical feedbacks, the effect of land cover changes on local and regional climates, and the differentiation of the role of biodiversity on climate trends and extremes. Improving the understanding of these processes involves integrating satellite observations and ground-based data, and it is necessary to refine their representation in climate models, aiming for higher spatial and temporal resolutions. Moreover, the inclusion of machine learning in hybrid modelling approaches, and the use of causal inference algorithms in combination with physical and AI modelling, offers vast potential to increase understanding of these complex interactions. Ultimately, understanding and predicting the feedback loops between vegetation diversity, ecosystem resilience, and climate stability is crucial for maintaining biodiversity and the health of our global ecosystems, and should remain a research priority.

## Acknowledgments

DGM acknowledges support from the European Research Council (ERC) through the HEAT Consolidator grant (101088405).

## Author contributions

DGM led the analysis and writing of the first draft of the review. All authors contributed to the discussions and the writing and editing of the manuscript.

## Competing interest statement

The authors declare no competing interests.

## References

Antão, L.H., Weigel, B., Strona, G., Hällfors, M., Kaarlejärvi, E., Dallas, T., Opedal, Ø.H., Heliölä, J., Henttonen, H., Huitu, O. and Korpimäki, E., 2022. Climate change reshuffles northern species within their niches. Nature Climate Change, 12(6), 587-592.
Albertson, J.D. and Parlange, M.B., 1999. Natural integration of scalar fluxes from complex terrain. Adv. Water Resour., 23(3): 239-252.
Armstrong, E., Valdes, P., House, J. and Singarayer, J., 2019. Investigating the feedbacks between CO2, vegetation and the AMOC in a coupled climate model. Clim. Dyn., 53(5-6): 2485-2500.
Arora, V.K. et al., 2020. Carbon concentration and carbon–climate feedbacks in CMIP6 models and their comparison to CMIP5 models. Biogeosciences, 17(16): 4173-4222.

Barriopedro, D., Garcia-Herrera, R., Ordonez, C., Miralles, D.G. and Salcedo, Sanz, S., 2023. Heat Waves: Physical Understanding and Scientific Challenges. Rev. Geophys., 61, e2022RG000780.

Belušić, D., Fuentes-Franco, R., Strandberg, G. and Jukimenko, A., 2019. Afforestation reduces cyclone intensity and precipitation extremes over Europe. Environ. Res. Lett., 14(7): 074009.

Bonan G., 2015: Ecosystems and Climate. In: Ecological Climatology: Concepts and Applications. Cambridge University Press; 2015:1-20.

Bonan, G.B., Patton, E.G., Finnigan, J.J., Baldocchi, D.D. and Harman, I.N., 2021. Moving beyond the incorrect but useful paradigm: reevaluating big-leaf and multilayer plant canopies to model biosphere-atmosphere fluxes – a review. Agric. For. Meteorol., 306: 108435.

Bonan, G.B., Lucier, O., Coen, D.R., Foster, A.C., Shuman, J.K., Laguë, M.M., Swann, A.L., Lombardozzi, D.L., Wieder, W.R., Dahlin, K.M. and Rocha, A.V., 2024. Reimagining Earth in the Earth system. Journal of Advances in Modeling Earth Systems, 16(8), p.e2023MS00401

Bowen, I.S., 1926. The Ratio of Heat Losses by Conduction and by Evaporation from any Water Surface. Physical Review, 27(6): 779-787.

Canadell, J.G., P.M.S. Monteiro, M.H. Costa, L. Cotrim da Cunha, P.M. Cox, A.V. Eliseev, S. Henson, M. Ishii, S. Jaccard, C. Koven, A. Lohila, P.K. Patra, S. Piao, J. Rogelj, S. Syampungani, S. Zaehle, and K. Zickfeld, 2021: Global Carbon and other Biogeochemical Cycles and Feedbacks. In Climate Change 2021: The Physical Science Basis. Contribution of Working Group I to the Sixth Assessment Report of the Intergovernmental Panel on Climate Change [Masson-Delmotte, V., P. Zhai, A. Pirani, S.L. Connors, C. Péan, S. Berger, N. Caud, Y. Chen, L. Goldfarb, M.I. Gomis, M. Huang, K. Leitzell, E. Lonnoy, J.B.R. Matthews, T.K. Maycock, T. Waterfield, O. Yelekçi, R. Yu, and B. Zhou (eds.)]. Cambridge University Press, Cambridge, United Kingdom and New York, NY, USA, pp. 673–816, doi:10.1017/9781009157896.007.

Casans, A. et al., 2023. Cloud condensation nuclei activation properties of Mediterranean pollen types considering organic chemical composition and surface tension effects. Atmos. Environ., 310: 119961.

Charney, J.G., 1975. Dynamics of deserts and drought in the Sahel. Quart. J. Roy. Meteorol. Soc., 101(428): 193-202.

Chen, C. et al., 2020. Biophysical impacts of Earth greening largely controlled by aerodynamic resistance. Science Advances, 6(47): eabb1981.

Cui, J. et al., 2020. Vegetation forcing modulates global land monsoon and water resources in a $CO_2$-enriched climate. Nature Communications, 11(1): 5184.

Deardorff, J.W., 1970. A numerical study of three-dimensional turbulent channel flow at large Reynolds numbers. J. Fluid Mech., 41(2): 453-480.

Díaz, S., Kattge, J., Cornelissen, J.H., Wright, I.J., Lavorel, S., Dray, S., Reu, B., Kleyer, M., Wirth, C., Colin Prentice, I. and Garnier, E., 2016. The global spectrum of plant form and function. Nature, 529(7585), pp.167-171.

Dickinson, R. E., Henderson-Sellers, A., Kennedy, P. J., and Wilson, M.F., 1986. Biosphere–Atmosphere Transfer Scheme (BATS) for the NCAR Community Climate Model, Technical Note NCAR/TN-275+STR. Boulder, Colorado: National Center for Atmospheric Research.

Dupont, S., M. R. Irvine, and C. Bidot, 2024: Morning Transition of the Coupled Vegetation Canopy and Atmospheric Boundary Layer Turbulence according to the Wind Intensity. J. Atmos. Sci., 81, 1225–1249, https://doi.org/10.1175/JAS-D-23-0201.1.

Durand, M., Murchie, E. H., Lindfors, A. V., Urban, O., Aphalo, P. J., and Robson, T. M., 2021. Diffuse solar radiation and canopy photosynthesis in a changing environment. Agric. For. Meteorol. 311, 108684.

Eltahir, E.A.B. and Bras, R.L., 1996. Precipitation recycling. Rev. Geophys., 34(3): 367-378.

Essl, F., Lenzner, B., Bacher, S., Bailey, S., Capinha, C., Daehler, C., ... and Roura-Pascual, N., 2020. Drivers of future alien species impacts: An expert-based assessment. Global Change Biology, 26(9), 4880-4893.

Eyring, V., N.P. Gillett, K.M. Achuta Rao, R. Barimalala, M. Barreiro Parrillo, N. Bellouin, C. Cassou, P.J. Durack, Y. Kosaka, S. McGregor, S. Min, O. Morgenstern, and Y. Sun, 2021: Human Influence on the Climate System. In Climate Change 2021: The Physical Science Basis. Contribution of Working Group I to the Sixth Assessment Report of the Intergovernmental Panel on Climate Change[Masson-Delmotte, V., P. Zhai, A. Pirani, S.L. Connors, C. Péan, S. Berger, N. Caud, Y. Chen, L. Goldfarb, M.I. Gomis, M. Huang, K. Leitzell, E. Lonnoy, J.B.R. Matthews, T.K. Maycock,

T. Waterfield, O. Yelekçi, R. Yu, and B. Zhou (eds.)]. Cambridge University Press, Cambridge, United Kingdom and New York, NY, USA, pp. 423–552, doi:10.1017/9781009157896.005.

Fan, X. et al., 2023. Escalating Hot-Dry Extremes Amplify Compound Fire Weather Risk. Earth's Future, 11, e2023EF003976.

Forster, P., T. Storelvmo, K. Armour, W. Collins, J.-L. Dufresne, D. Frame, D.J. Lunt, T. Mauritsen, M.D. Palmer, M. Watanabe, M. Wild, and H. Zhang, 2021: The Earth's Energy Budget, Climate Feedbacks, and Climate Sensitivity. In Climate Change 2021: The Physical Science Basis. Contribution of Working Group I to the Sixth Assessment Report of the Intergovernmental Panel on Climate Change [Masson-Delmotte, V., P. Zhai, A. Pirani, S.L. Connors, C. Péan, S. Berger, N. Caud, Y. Chen, L. Goldfarb, M.I. Gomis, M. Huang, K. Leitzell, E. Lonnoy, J.B.R. Matthews, T.K. Maycock, T. Waterfield, O. Yelekçi, R. Yu, and B. Zhou (eds.)]. Cambridge University Press, Cambridge, United Kingdom and New York, NY, USA, pp. 923–1054, doi:10.1017/9781009157896.009.

Forzieri, G., Alkama, R., Miralles, D.G. and Cescatti, A., 2017. Satellites reveal contrasting responses of regional climate to the widespread greening of Earth. Science, 356(6343): 1180-1184.

Friedlingstein, P. et al., 2023. Global Carbon Budget 2023. Earth Syst. Sci. Data, 15(12): 5301-5369.

Garratt, J.R., 1994. The Atmospheric Boundary-Layer - Review. Earth-Sci. Rev., 37(1-2): 89-134.

Hertog, S.J.D. et al., 2024. Effects of idealized land cover and land management changes on the atmospheric water cycle. Earth Syst. Dynam., 15(2): 265-291.

Huang, M. et al., 2019. Air temperature optima of vegetation productivity across global biomes. Nat. Ecol. Evol., 3(5):772-779, doi: 10.1038/s41559-019-0838-x.

Humboldt, A. von, Otté, E. C., and Bohn, H. G. (1850). Views of Nature: Or Contemplations on the Sublime Phenomena of Creation; with Scientific Illustrations. London: H. G. Bohn.

Insua-Costa, D., Senande-Rivera, M., Llasat, M.C. and Miguez-Macho, G., 2022. The central role of forests in the 2021 European floods. Environ. Res. Lett., 17(6): 064053.

Jacobs, A.F.G., Heusinkveld, B.G. and Holtslag, A.A.M., 2008. Towards Closing the Surface Energy Budget of a Mid-latitude Grassland. Boundary-Layer Meteorol., 126(1): 125-136.

Jaramillo, A., Mesa, O.J. and Raymond, D.J., 2018. Is Condensation-Induced Atmospheric Dynamics a New Theory of the Origin of the Winds? J. Atmos. Sci., 75(10): 3305-3312.

Jia, G., E. Shevliakova, P. Artaxo, N. De Noblet-Ducoudré, R. Houghton, J. House, K. Kitajima, C. Lennard, A. Popp, A. Sirin, R. Sukumar, L. Verchot, 2019: Land–climate interactions. In: Climate Change and Land: an IPCC special report on climate change, desertification, land degradation, sustainable land management, food security, and greenhouse gas fluxes in terrestrial ecosystems [P.R. Shukla, J. Skea, E. Calvo Buendia, V. Masson-Delmotte, H.-O. Pörtner, D.C. Roberts, P. Zhai, R. Slade, S. Connors, R. van Diemen, M. Ferrat, E. Haughey, S. Luz, S. Neogi, M. Pathak, J. Petzold, J. Portugal Pereira, P. Vyas, E. Huntley, K. Kissick, M. Belkacemi, J. Malley, (eds.)]

Joutsensaari, J. et al., 2015. Biotic stress accelerates formation of climate-relevant aerosols in boreal forests. Atmos. Chem. Phys., 15(21): 12139-12157.

Koppa, A., Keune, J., Schumacher, D. L., Michaelides, K., Singer, M., Seneviratne, S. I., and Miralles, D. G. 2024. Dryland self-expansion enabled by land–atmosphere feedbacks. Science 385, 967–972.

Köppen, W. (1936) Das geographische System der Klimate. Handbuch der Klimatologie (ed. by W. Köppen and R. Geiger), Vol 1 Part C pp. 1-44. Verlag von Gebrüder Borntraeger, Berlin.

Lemordant, L., Gentine, P., Stéfanon, M., Drobinski, P. and Fatichi, S., 2016. Modification of land-atmosphere interactions by CO2 effects: Implications for summer dryness and heat wave amplitude. Geophys. Res. Lett., 43(19): 10,240-10,248.

Lian, X. et al., 2022. Biophysical impacts of northern vegetation changes on seasonal warming patterns. Nature Communications, 13(1): 3925.

Liu, Y., Goodrick, S. & Heilman, W. Wildland fire emissions, carbon, and climate: Wildfire–climate interactions. For. Ecol. Manag. 317, 80–96 (2014).

Loreau, M., Barbier, M., Filotas, E., Gravel, D., Isbell, F., Miller, S. J., et al. (2021). Biodiversity as insurance: From concept to measurement and application. Biological Reviews, 96(5), 2333–2354.

Lorenz, R., Davin, E.L., Lawrence, D.M., Stckli, R. and Seneviratne, S.I., 2013. How Important is Vegetation Phenology for European Climate and Heat Waves? J. Clim., 26(24): 10077-10100.

Mahecha, M. D., Bastos, A., Bohn, F. J., Eisenhauer, N., Feilhauer, H., Hartmann, H., Hickler, T., Kalesse-Los, H., Migliavacca, M., Otto, F. E. L., Peng, J., Quaas, J., Tegen, I., Weigelt, A., Wendisch, M., and Wirth, C. 2022. Biodiversity loss and climate extremes — study the feedbacks. Nature 612, 30–32

Mahecha, M. D., Bastos, A., Bohn, F. J., Eisenhauer, N., Feilhauer, H., Hickler, T., ... and Quaas, J., 2024. Biodiversity and climate extremes: known interactions and research gaps. Earth's Future, 12(6), e2023EF003963.
Makarieva, A.M. and Gorshkov, V.G., 2007. Biotic pump of atmospheric moisture as driver of the hydrological cycle on land. Hydrol. Earth Syst. Sci., 11(2): 1013-1033.
Marvel, K. and Bonfils, C., 2013. Identifying external influences on global precipitation. Proc Natl Acad Sci U S A, 110(48): 19301-6.
McPherson, R.A., 2016. A review of vegetation—atmosphere interactions and their influences on mesoscale phenomena. Prog. Phys. Geog., 31(3): 261-285.
McVicar, T.R. et al., 2012. Global review and synthesis of trends in observed terrestrial near-surface wind speeds: Implications for evaporation. J. Hydrol., 416-417: 182-205.
Medlyn, B.E., Duursma, R.A., Eamus, D., Ellsworth, D.S., Prentice, I.C., Barton, C.V.M., Crous, K.Y., De Angelis, P., Freeman, M. and Wingate, L. (2011), Reconciling the optimal and empirical approaches to modelling stomatal conductance. Global Change Biology, 17: 2134-2144.
Meesters, A.G.C.A., Dolman, A.J. and Bruijnzeel, L.A., 2009. Comment on "Biotic pump of atmospheric moisture as driver of the hydrological cycle on land" by A. M. Makarieva and V. G. Gorshkov, Hydrol. Earth Syst. Sci., 11, 1033, 2007. Hydrol. Earth Syst. Sci., 13(7): 1299-1305.
Meidner, H. and Mansfield, T. A., 1968. Physiology of stomata. 179, London: McGraw-Hill.
Migliavacca, M., Musavi, T., Mahecha, M.D., Nelson, J.A., Knauer, J., Baldocchi, D.D., Perez-Priego, O., Christiansen, R., Peters, J., Anderson, K. and Bahn, M., 2021. The three major axes of terrestrial ecosystem function. Nature, 598(7881), 468-472.
Miralles, D.G., Gentine, P., Seneviratne, S.I. and Teuling, A.J., 2019. Land-atmospheric feedbacks during droughts and heatwaves: state of the science and current challenges. Ann N Y Acad Sci. 2019 Jan;1436(1):19–35. doi: 10.1111/nyas.13912.
Miralles, D.G. et al., 2016. Contribution of water-limited ecoregions to their own supply of rainfall. Environ. Res. Lett., 11(12): 1-12.
Monin, A. and Obukhov, A.M., 1954. Basic laws of turbulent mixing in the surface layer of the atmosphere. Trudy Geofiz. Inst. Acad. Nauk SSSR, 24(151): 163-187.
Mostamandi et al., 2022. Sea Breeze Geoengineering to Increase Rainfall over the Arabian Red Sea Coastal Plains. J. Hydrometeorol., 23(1): 3-24.
Musavi, T., Mahecha, M.D., Migliavacca, M., Reichstein, M., van de Weg, M.J., van Bodegom, P.M., Bahn, M., Wirth, C., Reich, P.B., Schrodt, F. and Kattge, J., 2015. The imprint of plants on ecosystem functioning: A data-driven approach. International Journal of Applied Earth Observation and Geoinformation, 43, 119-131.
Netherer, S., Lehmanski, L., Bachlehner, A., Rosner, S., Savi, T., Schmidt, A., ... & Gershenzon, J. 2024. Drought increases Norway spruce susceptibility to the Eurasian spruce bark beetle and its associated fungi. New Phytologist, 242(3), 1000-1017.
Orth, R., O, S., Zscheischler, J., Mahecha, M.D. and Reichstein, M., 2022. Contrasting biophysical and societal impacts of hydro-meteorological extremes. Environ. Res. Lett., 17(1): 014044.
Pedruzo-Bagazgoitia, X., Jimenez, P.A., Dudhia, J. and de Arellano, J.V.-G., 2019. Shallow cumulus representation and its interaction with radiation and surface at the convection grey zone Shallow cumulus representation and its interaction with radiation and surface at the convection grey zone. Mon. Weather Rev., 147(7): 2467-2483.
Peñuelas, J. et al., 2017. Shifting from a fertilization-dominated to a warming-dominated period. Nat Ecol Evol, 1(10): 1438-1445.
Piao, S. et al., 2020. Characteristics, drivers and feedbacks of global greening. Nat Rev Earth Environ, 1(1): 14-27.
Pongratz, J. et al., 2021. Land Use Effects on Climate: Current State, Recent Progress, and Emerging Topics. Curr Clim Change Reports, 7(4): 99-120.
Post, D.A. et al., 2014. Decrease in southeastern Australian water availability linked to ongoing Hadley cell expansion. Earth's Future, 2(4): 231-238.
Pörtner, H. O., Scholes, R. J., Arneth, A., Barnes, D. K. A., Burrows, M. T., Diamond, S. E., ... & Val, A. L. (2023). Overcoming the coupled climate and biodiversity crises and their societal impacts. Science, 380(6642), eabl4881.
Reichstein, M., Bahn, M., Mahecha, M.D., Kattge, J. and Baldocchi, D.D., 2014. Linking plant and ecosystem functional biogeography. Proceedings of the national academy of sciences, 111(38).13697-13702.

Roderick, M.L., Rotstayn, L.D., Farquhar, G.D. and Hobbins, M.T., 2007. On the attribution of changing pan evaporation. Geophys. Res. Lett., 34(17): L17403.
Schumacher, D.L., Keune, J., Dirmeyer, P. and Miralles, D.G., 2022. Drought self-propagation in drylands due to land–atmosphere feedbacks. Nat. Geosci., 15, 262–268.
Schumacher, D.L. et al., 2019. Amplification of mega-heatwaves through heat torrents fuelled by upwind drought. Nat. Geosci., 12, 712–717.
Seidel, D.J., Fu, Q., Randel, W.J. and Reichler, T.J., 2008. Widening of the tropical belt in a changing climate. Nat. Geosci., 1(1): 21–24.
Sellers, P. J., Mintz, Y., Sud, Y. C., and Dalcher, A., 1986. A simple biosphere model (SiB) for use within general circulation models. Journal of the Atmospheric Sciences, 43, 505–531.
Seneviratne S. I., Wartenburger, R., Guillod, B. P., Hirsch, A. L., Vogel, M. M., Brovkin, V., van Vuuren D. P., Schaller, N., Boysen, L., Calvin, K. V., Doelman, J., Greve, P., Havlik, P., Humpenöder, F., Krisztin, T., Mitchell, D., Popp, A., Riahi, K., Rogelj, J.,Schleussner, C.-F., Sillmann, J., and Stehfest, E. 2018a. Climate extremes, land–climate feedbacks and land-use forcing at 1.5°C, Phil. Trans. R. Soc. A., 376:20160450.
Seneviratne, S.I. et al., 2018b. Land radiative management as contributor to regional-scale climate adaptation and mitigation. Nat. Geosci., 11(2): 88–96.
Seneviratne, S.I., X. Zhang, M. Adnan, W. Badi, C. Dereczynski, A. Di Luca, S. Ghosh, I. Iskandar, J. Kossin, S. Lewis, F. Otto, I. Pinto, M. Satoh, S.M. Vicente-Serrano, M. Wehner, and B. Zhou, 2021: Weather and Climate Extreme Events in a Changing Climate. In Climate Change 2021: The Physical Science Basis. Contribution of Working Group I to the Sixth Assessment Report of the Intergovernmental Panel on Climate Change [Masson-Delmotte, V., P. Zhai, A. Pirani, S.L. Connors, C. Péan, S. Berger, N. Caud, Y. Chen, L. Goldfarb, M.I. Gomis, M. Huang, K. Leitzell, E. Lonnoy, J.B.R. Matthews, T.K. Maycock, T. Waterfield, O. Yelekçi, R. Yu, and B. Zhou (eds.)]. Cambridge University Press, Cambridge, United Kingdom and New York, NY, USA, pp. 1513–1766, doi:10.1017/9781009157896.013.
Shin, S.-H., Chung, I.-U. and Kim, H.-J., 2012. Relationship between the expansion of drylands and the intensification of Hadley circulation during the late twentieth century. Meteorol. Atmos. Phys., 118(3-4): 117-128.
Sikma, M. and Vilà-Guerau de Arellano, J., 2019. Substantial Reductions in Cloud Cover and Moisture Transport by Dynamic Plant Responses. Geophys. Res. Lett., 46(3): 1870-1878.
Sippel, S., Forkel, M., Rammig, A., Thonicke, K., Flach, M., Heimann, M., Otto, F.E., Reichstein, M. and Mahecha, M.D., 2017. Contrasting and interacting changes in simulated spring and summer carbon cycle extremes in European ecosystems. Environmental Research Letters, 12(7), p.075006.
Song, H. and Zhang, M., 2007. Changes of the Boreal Winter Hadley Circulation in the NCEP–NCAR and ECMWF Reanalyses: A Comparative Study. J. Clim., 20(20): 5191-5200.
Sporre, M.K., Blichner, S.M., Karset, I.H.H., Makkonen, R. and Berntsen, T.K., 2019. BVOC–aerosol–climate feedbacks investigated using NorESM. Atmos. Chem. Phys., 19(7): 4763-4782.
Stéfanon, M., Drobinski, P., D'Andrea, F. and de Noblet-Ducoudré, N., 2012. Effects of interactive vegetation phenology on the 2003 summer heat waves. J. Geophys. Res. Atmos., 117(D24), doi: 10.1029/2012JD018187.
Stocker, B.D. et al., 2013. Multiple greenhouse-gas feedbacks from the land biosphere under future climate change scenarios. Nat. Clim. Change, 3(7): 666-672.
Stocker, B.D. et al., 2019. Drought impacts on terrestrial primary production underestimated by satellite monitoring. Nat. Geosci., 12, 264–270.
Taylor, C.M., de Jeu, R.A., Guichard, F., Harris, P.P. and Dorigo, W.A., 2012. Afternoon rain more likely over drier soils. Nature, 489(7416): 423-6.
Teuling, A.J. et al., 2010. Contrasting response of European forest and grassland energy exchange to heatwaves. Nature Geosci., 3(10): 722-727.
Teuling, A.J. et al., 2017. Observational evidence for cloud cover enhancement over western European forests. Nat Commun, 8: 14065.
Thiery, W. et al., 2020. Warming of hot extremes alleviated by expanding irrigation. Nature Communications, 11(1): 290.
van Dijk, A.I.J.M. et al., 2015. Rainfall interception and the coupled surface water and energy balance. Agric. For. Meteorol., 214-215: 402-415.
van Heerwaarden, C.C., Vilà-Guerau de Arellano, J., Moene, A.F. and Holtslag, A.A.M., 2009. Interactions between dry-air entrainment, surface evaporation and convective boundary-layer development. Quart. J. Roy. Meteorol. Soc., 135(642): 1277-1291.

Vautard, R., Cattiaux, J., Yiou, P., Thépaut, J.-N. and Ciais, P., 2011. Northern Hemisphere atmospheric stilling partly attributed to an increase in surface roughness. Nat. Geosci., 3(11): 756-761.

Vilà-Guerau de Arellano, J., van Heerwaarden, C.C. and Lelieveld, J., 2012. Modelled suppression of boundary-layer clouds by plants in a $CO_2$-rich atmosphere. Nat. Geosci., 5(10): 701-704.

Vilà-Guerau de Arellano, J., Hartogensis, O., Benedict, I., Boer, H. de, Bosman, P. J. M., Botía, S., Cecchini, M. A., Faassen, K. A. P., González-Armas, R., Diepen, K. van, Heusinkveld, B. G., Janssens, M., Lobos-Roco, F., Luijkx, I. T., Machado, L. A. T., Mangan, M. R., Moene, A. F., Mol, W. B., Molen, M. van der, Moonen, R., Ouwersloot, H. G., Park, S., Pedruzo-Bagazgoitia, X., Röckmann, T., Adnew, G. A., Ronda, R., Sikma, M., Schulte, R., Stratum, B. J. H. van, Veerman, M. A., Zanten, M. C. van and van Heerwaarden, C. C., 2023. Advancing understanding of land–atmosphere interactions by breaking discipline and scale barriers. Ann. Ny. Acad. Sci. doi:10.1111/nyas.14956.

Vilà-Guerau de Arellano, J., Hartogensis, O. K., Boer, H. de, Moonen, R., González-Armas, R., Janssens, M., Adnew, G. A., Bonell-Fontás, D. J., Botía, S., Jones, S. P., Asperen, H. van, Komiya, S., Feiter, V. S. de, Rikkers, D., Haas, S. de, Machado, L. A. T., Dias-Junior, C. Q., Giovanelli-Haytzmann, G., Valenti, W. I. D., Figueiredo, R. C., Farias, C. S., Hall, D. H., Mendonça, A. C. S., Silva, F. A. G. da, Silva, J. L. M. da, Souza, R., Martins, G., Miller, J. N., Mol, W. B., Heusinkveld, B., Heerwaarden, C. C. van, D'Oliveira, F. A. F., Ferreira, R. R., Gotuzzo, R. A., Pugliese, G., Williams, J., Ringsdorf, A., Edtbauer, A., Quesada, C. A., Portela, B. T. T., Alves, E. G., Pöhlker, C., Trumbore, S., Lelieveld, J. and Röckmann, T., 2024. CloudRoots-Amazon22: Integrating Clouds with Photosynthesis by Crossing Scales. Bull. Am. Meteorol. Soc. 105, E1275–E1302.

Ward, P.J. et al., 2018. Dependence between high sea-level and high river discharge increases flood hazard in global deltas and estuaries. Environ. Res. Lett., 13(8): 084012.

Wever, N., 2012. Quantifying trends in surface roughness and the effect on surface wind speed observations. J. Geophys. Res. Atmos., 117(D11).

Willeit, M., Ganopolski, A. and Feulner, G., 2014. Asymmetry and uncertainties in biogeophysical climate–vegetation feedback over a range of $CO_2$ forcings. Biogeosciences, 11(1): 17-32.

Winkler, A.J., Myneni, R.B., Hannart, A., Sitch, S., Haverd, V., Lombardozzi, D., Arora, V.K., Pongratz, J., Nabel, J.E., Goll, D.S. and Kato, E., 2021. Slowdown of the greening trend in natural vegetation with further rise in atmospheric CO 2. Biogeosciences, 18(17): 4985-5010.

Yang, H., Huntingford, C., Wiltshire, A., Sitch, S. and Mercado, L., 2019. Compensatory climate effects link trends in global runoff to rising atmospheric $CO_2$ concentration. Environ. Res. Lett., 14(12): 124075.

Young, I.R., Zieger, S. and Babanin, A.V., 2011. Global Trends in Wind Speed and Wave Height. Science, 332(6028): 451-455.

Zeng, Z. et al., 2018a. Global terrestrial stilling: does Earth's greening play a role? Environ. Res. Lett., 13(12): 124013.

Zeng, Z. et al., 2018b. Impact of Earth greening on the terrestrial water cycle. J. Clim., 31(7): 2633-2650.

Zeng, Z. et al., 2019. A reversal in global terrestrial stilling and its implications for wind energy production. Nat. Clim. Change, 9(12): 979-985.

Zhang, Y., Piao, S., Sun, Y., Rogers, B.M., Li, X., Lian, X., Liu, Z., Chen, A. and Peñuelas, J., 2022. Future reversal of warming-enhanced vegetation productivity in the Northern Hemisphere. Nature Climate Change, 12(6): 581-586.

Zscheischler, J., Martius, O., Westra, S., Bevacqua, E., Raymond, C., Horton, R.M., van den Hurk, B., AghaKouchak, A., Jézéquel, A., Mahecha, M.D. and Maraun, D., 2020. A typology of compound weather and climate events. Nature reviews earth & environment, 1(7), pp.333-347.